\documentstyle[multicol,aps,prb,epsf]{revtex}
\begin{document}
\draft \title{On the possibility of a metallic phase in granular
  superconducting films} \author{Tai Kai Ng$^1$ and Derek K.K.
  Lee$^2$} \address{$^1$Department of Physics, Hong Kong University of
  Science and Technology, Clear Water Bay, Hong Kong} \address{$^2$Blackett
  Laboratory, Imperial College, Prince Consort Rd, London SW7 2BW,
  United Kingdom}
\date{ \today } \maketitle
\begin{abstract}
  We investigate the possibility of finding a zero-temperature
  metallic phase in granular superconducting films.  We are able to
  identify the breakdown of the conventional treatment of these
  systems as dissipative Bose systems. We do not find a
  metallic state at zero temperature. At finite temperatures, we find
  that the system exhibit crossover behaviour which may have
  implications for the analysis of experimental results.  We also
  investigate the effect of vortex dissipation in these systems.
\end{abstract} 

\pacs{74.76.-w,74.80.Bj,74.25.Dw}

\begin{multicols}{2}
\narrowtext

Recently, there has been renewed interests in the problem of
superconductor-insulator (SI) transition in low-$T_c$ thin films.
These systems undergo transition from superconductor to insulator as a
function of disorder, film thickness, or applied magnetic
field\cite{e1}. Theories\cite{t1,t2} describing this kind of
superconductor-insulator transition describe a second-order quantum
phase transition where a zero-temperature metallic phase exists as a
critical point between the superfluid and insulating phases. However,
recent experiments\cite{e2,e3} found that the metallic phase may be
more than a point in the phase diagram in certain systems. Instead,
the zero-temperature conductivity appears to be finite and non-zero in
a finite region in the phase diagram. It remains controversial whether
these systems remain metallic down to zero temperature. It has been
observed\cite{e4} that some of these systems become superconducting at
very low temperatures.

In this paper, we re-examine our theoretical understanding of the SI
transition\cite{t3,t4}. We will use a variational treatment which, in
principle, may describe superfluid, metallic and insulator phases at
zero temperature. We will discuss dissipation arising from normal
resistance or vortex motion. We find that, although there may be a
direct SI transition at zero temperature, finite-temperature crossover
phenomena may give rise to apparently metallic behaviour in
experiments. We also see that the two dissipation mechanisms affect
the low-temperature behaviour on different sides of the SI critical point.

\section{Introduction}

The destruction of the superconducting state at zero temperature is a
result of strong Coulomb interactions. Consider a lattice model for
Cooper pairs. Strong Coulomb repulsion leads to a Mott insulating
state where there is an integral number of Cooper pairs at each site.
However, if the system is coupled to a normal fluid, any excess charge
on a site (arising the motion of Cooper pairs from site to site) can
be screened to a certain extent by the normal component. This is
effective when the normal fluid has low resistance, $R_n$, because it
can respond rapidly to charge fluctuations. Since the coupling to the
normal fluid requires exchange of energy, the normal fluid can be
regarded as a dissipative environment for the Cooper pairs. The
strength of this dissipative coupling (or dynamic screening) is
inversely proportional to $R_n$.

We will also consider dissipation originating from the motion of
the normal cores in vortices\cite{bs,shim}. In this case, a similar picture
applies when we study the system in a dual representation where vortices
are the elementary bosonic objects.

In principle, dissipation may lead to non-superfluid but mobile Cooper
pairs (or vortices) at zero temperature\cite{ka}. To investigate this
issue, we require a formulation which can differentiate between the
superfluid, metallic and insulating states.  We will see below that we
can do so by considering separately local phase fluctuations which,
over a timescale of $\hbar/k_{\rm B}T$, are small compared to $2\pi$
and those which are larger than $2\pi$.  Previous work has
investigated either a superfluid-to-non-superfluid
transition\cite{chak86} or an insulator-to-conductor
transition\cite{fazio}. We want to see if these transitions are
\emph{separate} so that all three phases exist. Otherwise, they are
different descriptions of the \emph{same} critical point, in which
case the Bose metal does not exist in the model at zero temperature.
After establishing the ground state, we will also discuss the
finite-temperature behaviour of these systems.

\section{Dissipative Bose model}

For simplicity, we will consider first dissipation for the Cooper
pairs. Vortex dissipation will be discussed later. We will review the
conventional discussion of this problem and we extend previous treatments
by a more careful consideration of large phase fluctuations.

\begin{figure}[hbt]
  \begin{center} \leavevmode
    \epsfxsize=\columnwidth\epsfbox{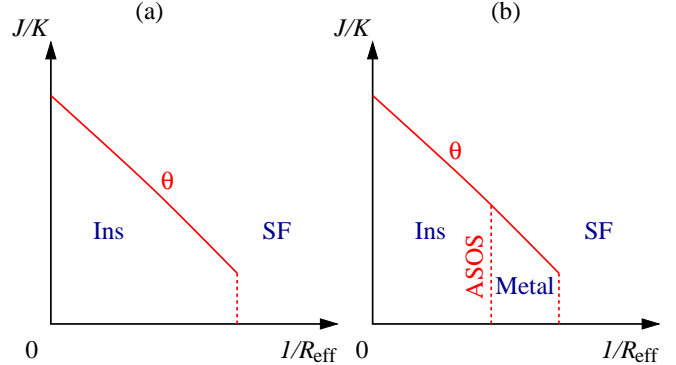} \caption{Possible
    scenarios at zero temperature. Ins: insulator, SF: superfluid.
    Solid and dashed lines denote first- and second-order transitions. ASOS
    line: transition for winding numbers in the absolute solid-on-solid
    model. $\theta$ line denotes transition for small fluctuations,
    $\theta$, as given in Ref.\ \protect\onlinecite{chak86}.}
    \label{fig:zerotemp} 
  \end{center}
\end{figure}

As our starting point, we use a model of superconducting grains on a
square array. We assume that well-defined Cooper pairs exist in each
grain so that we can treat them as charge-$2e$ bosons. An
imaginary-time action which describes the coupling between grains
is\cite{aes}:
\begin{equation}
\label{sboson}
S_{\rm boson}=\int^{\beta}_0 \!\!\!d\tau\left[{1\over 2K_b}
\sum_i (\dot\theta_i^{b})^{2} \! 
- J_b \sum_{i\nu}\cos\Delta_\nu\theta^b_i \right],
\end{equation}
where $\theta_i^{b}$ is the local superconducting phase of grain $i$,
and $\Delta_\nu\theta^b_i=\theta_{i+\nu}^b-\theta_i^b$. $J_b$ is the
Josephson coupling energy between nearest-neighbor grains. $K_b=
2e^2/C$ is the charging energy of a grain with self-capacitance $C$.
(We have set $\hbar=k_{\rm B}=1$, and $\beta=1/T$ is the inverse temperature.)
For large $J_b/K_b$, we expect a superconductor with long-range phase
coherence. When $J_b/K_b$ is small, however, the on-site repulsion
dominates and we have a Mott insulator. (Our calculations below will
focus on this limit.) The system becomes incompressible. (See vertical
axis on Fig.\ \ref{fig:zerotemp}.)  The phase, $\theta^b$, of the
local superconducting order parameter should fluctuate strongly at
each site due to the number-phase uncertainty relation.

We will now investigate the effect of dissipation on this bosonic Mott
transition.  We include dissipation phenomenologically. We assume that
an action of the Caldeira-Leggett kind\cite{cl,sz} is necessary so
that the charge currents ($\sim \Delta_\nu\theta^b$) will have ohmic
decay in the classical limit:
\begin{eqnarray}
\label{sdiss}
S_{\rm diss}=\frac{Q^2}{2}\sum_{i\nu}\int^{\beta}_0\!\!\!&&
d\tau\!\!\int^{\beta}_0\!\!\!d\tau' \alpha(\tau-\tau')\times \nonumber\\
&&\sin^2\left[\frac{\Delta_\nu\theta^b_i(\tau)-\Delta_\nu\theta^b_i(\tau')}{2Q}
\right],
\end{eqnarray}
where $\alpha(\tau)=(h/4e^2 R_n)[T/\sin(\pi{T}\tau)]^2$
and $Q=2$ reflecting the fact that the Cooper pairs has charge $2e$
while the dissipation is due to charge-$e$ electrons.

We will be interested in the destruction of superfluidity due to
enhanced phase fluctuations. As already mentioned, we have to be
careful about the compactness of the phase variables $\theta^b_i$. The
imaginary-time evolution of the phase can be separated into a periodic
part, $\theta_i$, and a non-periodic part\cite{t2}:
\begin{equation}
\label{instanton}
\theta_i^b(\tau)={2\pi{n_i}\tau\over\beta}+\theta_i(\tau)+\theta_{0i},
\end{equation}
where $\theta_i(\beta)=\theta_i(0)=0$. The boson action can be written as
\begin{eqnarray}
\label{scompact1}
S_{\rm boson} &=& {2\pi^2\over\beta K_b}\sum_i{n}_i^2
+ {1\over2K_b}\sum_i \int^{\beta}_0\dot{\theta}_i^2\,d\tau \nonumber\\
&-& J_b \sum_{i\nu} \int^{\beta}_0
        \cos\left(\Delta_\nu\theta_i(\tau)+
        {2\pi\tau\over\beta}\Delta_\nu n_i\right)d\tau,
\end{eqnarray}

To further simplify our calculation, we shall assume strong 
dissipation and keep only the $\Delta_\nu\theta_i$ terms in $S_{\rm diss}$
to second order. At low temperatures, we obtain\cite{fazio}:
\begin{eqnarray}
\label{scompact2}
S_{\rm diss} &&\rightarrow 
\frac{Q\pi}{4R_n}\sum_{i,\nu}|\Delta_\nu n_i|+
{1\over 8}\sum_{i\nu}\int^{\beta}_0\!\!\!d\tau\!\!\int^{\beta}_0\!\!\!d\tau'
\alpha(\tau-\tau') \times\nonumber\\
&&\cos\left[{2\pi(\tau-\tau')\over{Q\beta}}\Delta_\nu n_i\right]
\left[\Delta_\nu\theta_i(\tau)-\Delta_\nu\theta_i(\tau')\right]^2.
\end{eqnarray}

We can now discuss possible scenarios for the zero-temperature phase
diagram of the system. First of all, let us concentrate on the part of
the action which involves only the ``winding numbers'', $n_i$. We can
ignore the charging term in (\ref{scompact1}) proportional to $n_i^2$
because it vanishes as $T\rightarrow 0$.  The winding numbers are
controlled by the first term in (\ref{scompact2}).  This is, in fact,
the ``absolute solid-on-solid'' (ASOS) model which has a
``roughening'' transition (of the Kosterlitz-Thouless type) at $R_n =
R_c^{\rm ASOS}\simeq 0.6(h/Qe^2)$. For large $R_n$, the phase at each
site fluctuate wildly with little correlation between different sites.
This is what is expected (from number-phase uncertainty) in an
insulator where the local particle number does not fluctuate. For
small $R_n$, the system becomes ``smooth'' in the sense that large
excursions in the phase are suppressed. The system is now compressible
and the charges are mobile. This model has been used to describe an
insulator-conductor transition in normal tunneling junction networks
when $R_n$ is small enough \cite{fazio}.

We have seen that the Mott insulator breaks down and charges are mobile at
small $R_n$. What about superfluidity for these mobile charges? This
requires long-range phase coherence in the system. In other words, in
addition to a ``smooth'' $n$-field, the fluctuations of $\theta$ at
different sites must also be coherent. Therefore, in principle, we may
have a superfluid or metallic state for these mobile charges,
depending on whether the phase stiffness for $\theta$ fluctuations is
finite or not.

If we ignore the coupling of the $\theta$-field to the winding numbers
$n_i$, then we expect a superfluid at small $R_n$ at $T=0$.
($R_n<h/2e^2$ in two dimensions\cite{chak86}.) A primary purpose of
this paper is to investigate whether the onset of a finite phase
stiffness for $\theta$ coincides with the appearance of the smooth
phase in the SOS model for the winding number (\emph{i.e.} a direct
superfluid-insulator transition, as shown in
Fig.\ \ref{fig:zerotemp}a.) Another scenario is that a metal phase
exists for intermediate values of $R_n$ where the ASOS model is smooth
\emph{before} long-range phase coherence sets in at an even lower
value of $R_n$ (Fig.\ \ref{fig:zerotemp}b).

The actions (\ref{scompact1}) and (\ref{scompact2})
form the basis of our calculations. The model cannot
be solved exactly even without the dissipative term. We shall
pursue a variational approach since we are only
interested at the qualitative behaviour of the system ---
in particular, whether a zero-temperature metallic phase exists
under appropriate conditions. We consider the following trial action:
\begin{eqnarray}
\label{svar}
S_0&=& \sum_i\!\!\int^{\beta}_0\!\!\!d\tau\left[
\frac{1}{2K_b}\dot{\theta}_i^2
+{J_{\rm eff}\over2}\sum_{\nu}(\Delta_\nu\theta_i)^2\right]
\nonumber\\
+\frac{1}{8}& &\!\!\!\!\!\!\sum_{i\nu}
\int^{\beta}_0\!\!\!d\tau\!\!\int^{\beta}_0\!\!\!d\tau'\,
\alpha_{\rm eff}(\tau-\tau')
        \left[\Delta_\nu\theta_i(\tau)-\Delta_\nu\theta_i(\tau')\right]^2
\nonumber\\
+& &\!\!\!\!\!\!\frac{2\pi^2}{\beta{K}_b}\sum_i n_i^2+
\sum_{i\nu}\left[{{Q\pi}\over{4R}_{\rm eff}}|\Delta_\nu n_i|
                -\beta J_{\rm MS}\delta_{\Delta_\nu n_i}\right],
\end{eqnarray}
where $\alpha_{\rm eff}(\tau)/\alpha(\tau)=R_n/R_{\rm eff}$.  $J_{\rm
eff}$, $J_{\rm MS}$ and $R_{\rm eff}$ are parameters to be determined
variationally. Note that the solid-on-solid part of the model has been
modified by the presence of the $J_{\rm MS}$ term. Similar to the
other terms in the ASOS model, it also suppresses the spatial
fluctuations in the winding number. We therefore expect this modified
solid-on-solid (MSOS) model to be similar to the ASOS model with a
shifted critical point $R^{\rm MSOS}_c$.

The possibilities of superconductor, insulator, and metal phases at
zero temperature are all included in $S_0$. A finite value for the
phase stiffness, $J_{\rm eff}$, indicates that we have a
superconductor (marked ``SF'' in Fig.\ \ref{fig:zerotemp}). 
If $J_{\rm eff}=J_{\rm MS}=0$, then the system is
non-superfluid. To determine whether it is an insulator or a metal, we
examine the large phase fluctuations, \emph{i.e.} the SOS model for
the winding numbers. The system is an insulator if $R_{\rm eff}>R^{\rm
MSOS}_c$ so that the SOS model is in the rough phase. If $R_{\rm
eff}<R^{\rm MSOS}_c$, the SOS model is in the smooth phase and we have
a metallic state. (See Fig.\ \ref{fig:zerotemp}b. Note that the ASOS
and MSOS models are the same if $J_{\rm eff}=J_{\rm MS}=0$.)

The variational parameters are determined by minimizing the free
energy per unit volume given approximately by $F=F_{\rm 0}+
\langle S_{\rm dual}+ S_{\rm diss}-S_0\rangle_0 /\beta{L^2}$,
where $F_0$ is the free energy
calculated using $S_0$ and $\langle\cdots\rangle_0$ 
denotes averages taken with respect to $S_0$.We
obtain the mean-field equations:

\begin{eqnarray}
\label{mfe}
R_{\rm eff} & = & R_n, \nonumber\\ 
J_{\rm MS} & = & J_b e^{-\langle|\Delta\theta|^2\rangle/2}, \nonumber\\ 
J_{\rm eff} & = & J_{\rm MS}P_{\rm SOS}(0),
\end{eqnarray}
where
\begin{equation}
\label{mf3} 
\langle |\Delta\theta|^2\rangle = \frac{1}{2\beta L^2}
\sum_{\vec{q},i\omega_n}\gamma(\vec{q})G_{0\theta}(\vec{q},i\omega_n),
\end{equation}
with $G^{-1}_{0\theta}(\vec{q},i\omega_n)= 
\omega_n^2/K_b+\gamma(\vec{q})(J_{\rm eff}+|\omega_n|/4R_{\rm eff})$. 
$\gamma(\vec{q})
=4[\sin^2(q_x/2)+\sin^2(q_y/2)]$ is the lattice dispersion relation.
$P_{\rm SOS}(m)=\langle\delta(
|\Delta_\nu n_i|-m)\rangle_{\rm MSOS}$ is the probability that the 
nearest-neighbor integer difference $|\Delta_\nu n_i|=m$ in the MSOS model.
Note that we can also regard our trial action as a ``Hartree'' 
decoupling of the fields $\theta_i$ and $n_i$.

For the small phase fluctuations, the critical point for the onset of
a finite $J_{\rm eff}$ is given by Chakravarty \emph{et al.}\cite{chak86}:
$R_c^{\theta} = h/Qe^2$, with $J_{\rm eff}$ becoming exponentially
small as $R$ approaches $R_c^{\theta}$:
\begin{equation}
\label{Jcritical}
\left(\frac{J_{\rm eff}}{J_b}\right) \sim 
\left(\frac{J_b R}{K_b}\right)^{R^{\theta}_c/(R^{\theta}_c - R_{\rm eff})}
\end{equation}

To determine the ground-state properties of the system, we also need to
examine the SOS sector of the model. We see that the $J_{\rm MS}$
term dominates the MSOS model at low temperatures, and so the SOS
sector is smooth whenever $J_{\rm eff}$ is finite. When $J_{\rm eff}$
vanishes, we find that $R_n$ is already above the critical value for
the SOS critical point (\emph{i.e.}, $R_c^{\theta} > R_c^{\rm ASOS}$).
Therefore, the winding-number sector is always rough when $J_{\rm
  eff}=0$ so that the system is an insulator. This means that, at the
level of this mean-field calculation, we cannot have a metallic phase
at zero temperature. We see that the system has only one quantum
critical point as we change $R_n$ (marked ``SI'' in
Fig.~\ref{fig:phase}): $R_c = R_c^{\theta} = R_c^{\rm MSOS}$. This
corresponds to the scenario in Fig.\ \ref{fig:zerotemp}a.

\section{Finite Temperature}

We will now discuss the system at finite temperature. We will see that
the winding-number fluctuations have important consequences for the
behaviour of the system because the MSOS model governing these
fluctuations has an apparent finite-temperature phase transition.
These effects show up as crossover behaviour as the system is cooled to
zero temperature.

\begin{figure}[hbt]
  \begin{center} \leavevmode
    \epsfxsize=\columnwidth\epsfbox{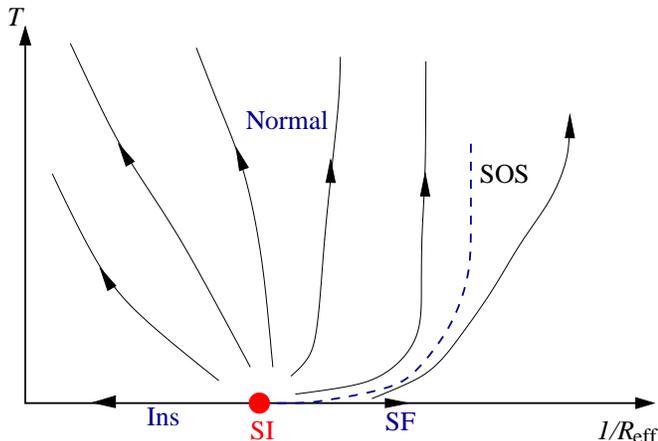} \caption{Phase
    diagram. Arrows indicate a schematic renormalization-group flow. A
    superfluid-insulator critical point (SI) separates the superfluid
    (SF) and insulating (Ins) phases at $T=0$. The system is normal at
    finite temperatures. Dashed line denotes the lines of (nominal)
    roughening transitions in the winding numbers: smooth phase at
    small $R_{\rm eff}$ and rough at large $R_{\rm eff}$. Our
    treatment is valid on the right of the dashed line.}
    \label{fig:phase} \end{center}
\end{figure}

To see this, we note that the critical value $R_c^{\rm MSOS}$ for
the transition in the MSOS model (\ref{svar}) depends on temperature.
At high temperatures, the $\beta J_{\rm MS}$ term in the MSOS model
becomes unimportant, and so we expect $R_c^{\rm MSOS}$ to decrease
towards $R_c^{\rm ASOS}=0.6(h/Qe^2)$ as the temperature increases.
This is indicated by dashed line in Fig.~\ref{fig:phase}. In other
words, for a resistance in the region $0.6Q < Q^2e^2R_n/h < 2$, the
system will cross a roughening transition for the winding numbers as
we increase the temperature.  Although this transition is probably an
artefact of the variational treatment, we believe that it will
manifest itself as a crossover phenomenon in the system.

More precisely, while there appears to be two correlation lengths in
this formulation ($\xi_\theta$ and $\xi_{\rm SOS}$ for the small and
large fluctuations respectively), there is only one true phase
correlation length, $\xi$. This should follow the shorter of
$\xi_\theta$ and $\xi_{\rm SOS}$. So, the SOS model does not give rise
to a true divergence in observable quantities, since $\xi_{\theta}$ is
finite at all finite temperatures. Instead, the divergence will be cut
off when $\xi_{\rm SOS}$ becomes comparable to $\xi_{\theta}$.

More generally, we expect physical quantities, such as the
conductivity, to depend on both temperature and the proximity of the
resistance to the critical value, $R_n-R_c$. For instance, in the
smooth phase of the SOS model (to the right of the dashed line in
Fig.\ \ref{fig:phase}), the correlation length $\xi_{\rm SOS}$ is
finite for fluctuations of the winding numbers about a smooth
background.  This affects dynamic quantities such as the
conductivity which should therefore depend on \emph{both} $R_n-R_c$ and $T$.

On the rough side of the SOS line, we expect no long-range order in
the winding number. The conductivity may not exhibit signs of
superfluidity. In fact, it may appear metallic or even insulating,
\emph{even at superfluid values of $R_n$}, as long as we are looking at
temperatures above the temperature, $T_{\rm SOS}$, where we cross the SOS
transition line. The temperature scale for this SOS crossover is given
by $J_{\rm eff}$. This can become very small close to the quantum
critical point (see eq.\ (\ref{Jcritical})) or in strongly disordered
systems.  We see that the true critical behaviour of the
superfluid-insulator transition is hard to access experimentally.

This discussion warns us that, unless we work at extremely low
temperatures, the critical behaviour of the system may not follow a
simple one-parameter scaling scheme (when $R_n$ is close to $R_c$ so
that the system is to the left of the SOS line in Fig.\ 
\ref{fig:phase}). We believe that this may be an important source of
difficulties for the scaling analysis of experimental data, and may be
responsible for the observation of an apparent metallic phase in some
experiments\cite{e2,e3}.

To be cautious, we should stress that this result depends on the
observation that $R_c^{\theta} > R_c^{\rm ASOS}$ (see discussion below
eq.\ (\ref{Jcritical})) so that it is sensitive to our estimates of
$R_c^{\theta}$ and $R_{\rm eff}$. For instance, we note that $R_n$ is
unrenormalized in our variational equations (\ref{mfe}). A more
careful treatment of the dissipative term might renormalize this
quantity and therefore shift the relative positions of the critical
points of the $\theta$ and SOS sectors. We will assume that these
estimates are correct in the next section.

The above analysis is based on a treatment which treats the coupling
between the small and large phase fluctuations ($\theta$ and $n$) in a
Hartree-like manner. In the next section, we will check that this is
reasonable by considering higher-order fluctuations. We will see that
the crossover effect mentioned above shows up as the breakdown of our
Hartree-like decoupling of the small and large phase fluctuations.

\section{Beyond Gaussian Fluctuations}

To consider higher-order fluctuations, let us examine the free energy
density $f$. This can be written as:
\begin{eqnarray}
f - f_0 &=& 
-\ln \langle \exp[-(S-S_0)]\rangle_0/\beta L^2 \nonumber\\
&\simeq& 
[\langle S-S_0\rangle_0 - \langle (S-S_0)^2\rangle_{c0}/2]/\beta L^2 + \cdots
\end{eqnarray}
where averages are taken with respect to the trial action $S_0$ and
$\langle A B\rangle_c = \langle A B\rangle -\langle A\rangle \langle
B\rangle$ denotes the connected part of the correlation function.
Minimizing the first term in this expansion gives the variational
treatment in the previous section. To consider the validity of this
approach, we should check that higher-order terms do not
diverge. These correspond to fluctuations beyond the Hartree-like
treatment in the previous section. We restrict our attention to the
first correction.

We can separate the Josephson and dissipative parts of $S - S_0$ as
$\delta S^{\rm J} + \delta S^{\rm D}$ where
\begin{eqnarray}
\delta S^{\rm J} = 
\sum_{i\nu} \int^{\beta}_0\!\!\!d\tau\Bigg[
        &-&J_b\cos\left(\Delta_\nu\theta_i+
        {2\pi\tau\over\beta}\Delta_\nu n_i\right) \nonumber\\
        & &- \frac{J_{\rm eff}}{2}(\Delta_\nu \theta_i)^2 
        + J_{\rm MS}\delta_{\Delta_\nu n_i}\Bigg]
\end{eqnarray}
\begin{eqnarray}
&&\delta S^{\rm D} = 
\sum_{i\nu}\int^{\beta}_0\!\!\!d\tau\!\!\!\int^{\beta}_0\!\!\!d\tau'
        \alpha(\tau-\tau')\times
\nonumber\\
&&\left[
\cos\left(\frac{2\pi(\tau-\tau')}{Q\beta}\Delta_\nu n_i\right)\!-\!1
\right]
\left[\Delta_\nu \theta_i(\tau)-\Delta_\nu \theta_i(\tau')\right]^2
\end{eqnarray}

As we approach the SI transition ($J_{\rm eff}\rightarrow 0$, $R_{\rm
eff}\rightarrow R_c$) at zero temperature, we find that the singular 
part of $\langle(\delta S^{\rm J})^2\rangle/\beta L^2$ 
comes from fluctuations in $\theta$, scaling as
$J_{\rm eff}^{2R_{\rm eff}/R_c - 1}$. We see that $\langle(\delta
S^{\rm J})^2\rangle$ does not diverge even at the critical point.  In
the non-superfluid phase ($J_{\rm eff}=0$), these fluctuations are
proportional to $T$ as $T\rightarrow 0$.

The contributions to $\langle\delta S^{\rm J} \delta S^{\rm
D}\rangle/\beta L^2$ are also finite, scaling as $T$ when
$T\rightarrow 0$ at finite $J_{\rm eff}$, and scaling as $J_{\rm eff}$ when
$J_{\rm eff}\rightarrow 0$ at finite $T$.

We find that the most singular term comes from $\langle(\delta
S^{\rm D})^2\rangle$. Let $A_{\tau,\tau'} = \sum_{i\nu} [\Delta_\nu
\theta_i(\tau)-\Delta_\nu \theta_i(\tau')]^2$ and $B_{\tau,\tau'} =
\sum_{i\nu}\alpha(\tau-\tau')[\cos(2\pi(\tau-\tau')\Delta_\nu n_i/Q\beta)-1]/2$.
Then, the contribution from
\begin{eqnarray}
I &=&\frac{1}{\beta L^2} \int \langle
A_{\tau_1,\tau_1'}\rangle\langle A_{\tau_2,\tau_2'} \rangle\langle
B_{\tau_1,\tau_1'}B_{\tau_2,\tau_2'}\rangle_c d\tau_1 d\tau_1'd\tau_2
d\tau_2'\nonumber\\
&\sim& \frac{1}{\beta}\sum_{\omega\omega'}
\frac{\sum_{{\bf r}\mu\nu}
\langle g_{\mu}({\bf r},\omega) g_{\nu}({\bf 0},\omega')\rangle_c}
{(4J_{\rm eff}R_{\rm eff}+|\omega|)(4J_{\rm eff}R_{\rm eff}+|\omega'|)}\\
&=& \sum_{\omega>2\pi T|\Delta_\nu n_{\bf r}|\atop
        \omega'>2\pi T|\Delta_\mu n_0|}\frac{T^3\sum_{{\bf r}\mu\nu}
\langle |\Delta_\nu n_i||\Delta_\mu n_0|\rangle_c}
{(4J_{\rm eff}R_{\rm eff}+|\omega|)(4J_{\rm eff}R_{\rm eff}+|\omega'|)}
\end{eqnarray}
where $g_{\nu}({\bf r}\!\!=\!\!{\bf r}_i,\omega)= {\rm
min}(\omega,2\pi Q^{-1} T|\Delta_\nu n_i|)$. The numerator is a connected
correlation function for the MSOS model. We expect it to have
exponential decay with correlation length $\xi_{\rm SOS}$ in the
smooth phase (and power-law decay in the rough phase).

In the smooth phase of the MSOS model where $J_{\rm eff}$ is also
finite, we see that $I \sim T \xi_{\rm SOS}^2 \ln (K_b/J_{\rm eff})$
as $T\rightarrow 0$.  On the other hand, we expect the quantity
$\sum_{ij\mu\nu}\langle|\Delta_\nu n_i||\Delta_\mu n_j|\rangle_c$ to
have the same critical behaviour as the energy fluctuations --- it
diverges as we cross the line of SOS critical points.  As discussed in
the previous section, this will not be a true divergence, but only a
crossover. Nevertheless, this means that this contribution from
$\langle(\delta S^{\rm D})^2\rangle$ will be large if we cross the SOS
transition line as we raise the temperature in the superfluid phase.

This marks the breakdown of our treatment of the phase fluctuations in
this model (in the region to the left of the dashed line in
Fig.~\ref{fig:phase}). However, since no divergences occur if we work
at zero temperature, the conclusion of a direct superfluid-insulator
transition appears robust (subject to the remarks at the end of the
previous section about the accuracy of our estimates of the relative
values for the critical points for the two sectors of the model.)

\section{Vortex Dissipation}

We will now discuss dissipation by vortex motion. Microscopically, this
is due to the motion of the normal vortex core. We will, however, follow a
phenomenological approach here. 

For this purpose, it is convenient to study the system in a vortex
representation. Fluctuations can be described by vortex loops in
Euclidean space-time. In particular, the superfluid state for the
Cooper pairs corresponds to a vortex insulator\cite{t1} where there is
a gap to the addition of a vortex --- the Meissner effect. Conversely,
the duality transformation shows that the Meissner phase of the
vortices correspond to an insulating state for Cooper pairs
(\emph{i.e.}, there is a gap to density excitations.)

To obtain the vortex representation, a duality transformation can be
applied to the action \ (\ref{sboson}) to obtain the dual action
$S_{\rm dual}=S_A+S_{\rm v}$ for vortices, where\cite{dual}
\begin{eqnarray}
\label{sdual}
S_A & = & \sum_i\int^{\beta}_0\!\!\!d\tau
\left[ \frac{1}{2J_b} |(\nabla\times\vec{A})^s|_i^2 + 
K_b |(\nabla\times\vec{A})^{\tau}|_i^2\right], \nonumber\\
S_{\rm v} & = & \int^{\beta}_0\!\!\!d\tau\!
\left[\frac{1}{2K_v}\sum_i(A^{\tau}_i
-\dot{\theta}^v_i)^2-J_v\sum_{i\nu}\cos D_\nu\theta_i^v\right]
\end{eqnarray}
where $D_\nu\theta_i=\Delta_\nu\theta_i-A_i^\nu$ is a covariant
derivative. The internal gauge field, $A$, is defined so that
$\nabla\times\vec{A}$ is the boson 3-current. Its action, $S_A$,
describes the phonons in the (original) boson superfluid. (The
superscripts $s$ and $\tau$ denote the spatial and temporal components
respectively.) The action $S_{\rm v}$ describes vortices in the
system: $\theta^v_i$ is the phase of the {\em vortex} wavefunction on
site $i$ of the dual lattice. We have introduced the terms $K_v \sim
J_b$ and $J_v\sim 2e\sqrt{J_b/c}$ to characterize the core energy and
the hopping integral of the vortices respectively\cite{dual}. The
coupling of the vortex phase to the gauge field expresses the fact
that vortices are advected by the current of the original bosons.

The qualitative behaviour of the system should not depend on details
of the vortex interaction as long as it is short-ranged. We therefore
choose the lattice spacing, $d$, for the dual model to be of the order
of the penetration depth (of the original Cooper pairs), and include
only on-site repulsion for vortices. For simplicity, we choose a
square lattice.


As with the boson model discussed above, we expect the system to have
a superfluid-insulator transition as we increase $K_v/J_v$. To include
dissipation phenomenologically, we again assume that an action of the
Caldeira-Leggett kind\cite{cl,sz} so that the vortex currents ($\sim
D_\nu\theta^v$) will decay with a decay rate proportional to the
current:
\begin{eqnarray}
\label{sdissv}
S_{\rm diss,v}=\frac{1}{2}\sum_{i\nu}
\int^{\beta}_0\!\!\! d\tau\!\!&&\int^{\beta}_0\!\!\!d\tau'
\ \alpha(\tau-\tau')\times \nonumber\\
&&\sin^2\left[\frac{D_\nu\theta^v_i(\tau)-D_\nu\theta^v_i(\tau')}{2}\right],
\end{eqnarray}
where $\alpha(\tau)=(h/4e^2 R_v)[T/\sin(\pi{T}\tau)]^2$ with
$4e^2R_v/h\sim (1-t)/t$ and $t \sim {e}^{-\eta{d}^2/\hbar}$ is the
tunneling resistance of vortices from one grain to another\cite{shim}.
Note that we have set $Q=1$ in this case because we do not have a
microscopic reason for the dissipative mechanism to involve objects
with a charge that is different from the bosons. The vortex viscous
drag coefficient $\eta$ is given by $\Phi_o H_{c2}/R_n c^2$ where
$R_n$ is the normal-state resistance of the superconductor, $H_{c2}$
is the upper critical field, and $\Phi_0=hc/2e$ is the flux quantum.
The details of the relationship between $R_v$ and $R_n$ are not
important here. It suffices to note that they are inversely related to
each other and comparable when both are of the order of $h/e^2$. The
coupling to the internal gauge field is required by gauge invariance.

We see that this model is similar to the one discussed in the previous
sections, except that the bosons are now coupled to an internal gauge
field. We can again separate the imaginary-time evolution of the phase
into a periodic part, $\theta_i(\tau)$, and a non-periodic
part, $2\pi n_i \tau/\beta$, (\ref{instanton}) to obtain:
\begin{eqnarray}
\label{scompact1v}
S_{\rm dual} &=& S_A +{2\pi^2\over\beta K_v}\sum_i{n}_i^2
+ {1\over2K_v}\sum_i \int^{\beta}_0\dot{\theta}_i^2\,d\tau \nonumber\\
&-& J_v \sum_{i\nu} \int^{\beta}_0
        \cos\left(D_\nu\theta_i(\tau)+
        {2\pi\tau\over\beta}\Delta_\nu n_i\right)d\tau,
\end{eqnarray}
Repeating the treatment in the previous sections (with $Q=1$), we have
the trial action:
\begin{eqnarray}
\label{svarv}
&&S_0= \sum_i\!\!\int^{\beta}_0\!\!\!d\tau\left[
\frac{1}{2K_v}\dot{\theta}_i^2
+{J_{\rm eff}\over2}\sum_{\nu}(D_\nu\theta_i)^2\right]
\nonumber\\
&+&\frac{1}{8}\sum_{i\nu}
\int^{\beta}_0\!\!\!d\tau\!\!\int^{\beta}_0\!\!\!d\tau'\,
\alpha_{\rm eff}(\tau-\tau')
        \left[D_\nu\theta_i(\tau)-D_\nu\theta_i(\tau')\right]^2
\nonumber\\
&+&\frac{2\pi^2}{\beta K_v}\sum_i n_i^2+
\sum_{i\nu}\left[{\pi\over{4R}_{\rm eff}}|\Delta_\nu n_i|
                -\beta J_{\rm MS}\delta_{\Delta_\nu n_i}\right],
\end{eqnarray}
The variational parameters are now given by $R_{\rm eff}=R_v$, $J_{\rm
eff} = J_{\rm MS} P_{\rm SOS}(0)$ and
\begin{eqnarray}
\label{mfev} 
J_{\rm MS} &=& J_v 
\exp\left[-(\langle|\Delta\theta|^2\rangle+\langle A\rangle)/2\right], 
\nonumber\\ 
\langle|\Delta\theta|^2\rangle & = & {1\over 2\beta{L}^2}
\sum_{\vec{q},i\omega_n}\gamma(\vec{q})G_{0\theta}(\vec{q},i\omega_n),
\\  \nonumber
\langle|A|^2\rangle & = &
{1\over 2\beta{L}^2}\sum_{\vec{q},i\omega_n}G_{0A}(\vec{q},i\omega_n),
\end{eqnarray}
where $G^{-1}_{0A}(\vec{q},i\omega_n)=\omega_n^2/J_b+ 2K_b\gamma(\vec{q})
+|\omega_n|/4{R}_{\rm eff}+J_{\rm eff}$ and
$G^{-1}_{0\theta}(\vec{q},i\omega_n)=
\omega_n^2/K_v+\gamma(\vec{q})(J_{\rm eff}+|\omega_n|/4R_{\rm eff})$. 

The main effect of the gauge fields is to reduce the vortex repulsion
$K$.  (For weak boson repulsion, $K_v \rightarrow K^* =
(K_v/4\pi^2)(J_b/2K_b)(h/4e^2R_v)$.) Since this is essentially a
high-energy cutoff for the physical effects we are considering, it
should not affect the critical point for the onset of a finite $J_{\rm
eff}$ for small vortex phase fluctuations. We therefore conclude that
this dissipative mechanism will also give rise to a direct
superfluid-to-insulator transition in the vortex liquid as $R_v$ is
increased.  Note that the vortex resistance, $R_v$ is large when the
resistance, $R_n$, of the normal fluid is low. Therefore, we have
qualitatively the same zero-temperature behaviour here as in the
previous model (for direct dissipation from Cooper pair motion) in
that the system, in terms of electrical transport by the original
Cooper pairs, is superfluid for small $R_n$ and insulating for large
$R_n$. The exact value of the critical resistance is more difficult to
extract as it depends in detail on the dependence of $R_v$ on
$R_n$. However, it can be verified that the critical point occurs when
$e^2R_n/h$ is of the order of unity.

At finite temperatures, we again expect crossover behaviour. This
applies to the \emph{insulating} side of the (original) SI transition,
whereas the crossover behaviour of the previous Cooper-pair model
affects the superfluid side of the transition. In terms of $R_n$
(instead of $R_v$), this model predicts that, for insulating values of
$R_n$ (\emph{i.e.}\ vortex superfluidity at $T=0$), the
finite-temperature conductivity for $R_n$ may appear metallic or even
exhibit signs of (charge) superfluidity above the crossover
temperature.

\section{Finite Magnetic Field}

Finally, we discuss the effect of a finite magnetic field, $B$.
This gives rise to a non-zero chemical potential, $\mu_v$, for
vortices so that there are a finite density of vortices in the ground
state: $\rho_v = B/\Phi_0$. Again, the movement of the MSOS crossover
as a function of vortex density (at fixed $R_n$ or $R_v$) will affect
the finite-temperature behaviour of the system, and hence the analysis
of the experimental data. 

Experimentally, we see that the system goes from superconducting to
insulating as we increase $B$. Some experiments\cite{e2,e3} indicate
that there may be a metallic phase between the superconducting and
insulating phases. However, there is also evidence\cite{e4} at low
applied fields that the metallic behaviour only occurs at intermediate
temperatures, and that the true zero-temperature phase may be a
superconductor after all.

\begin{figure}[hbt]
  \begin{center} \leavevmode
    \epsfxsize=\columnwidth\epsfbox{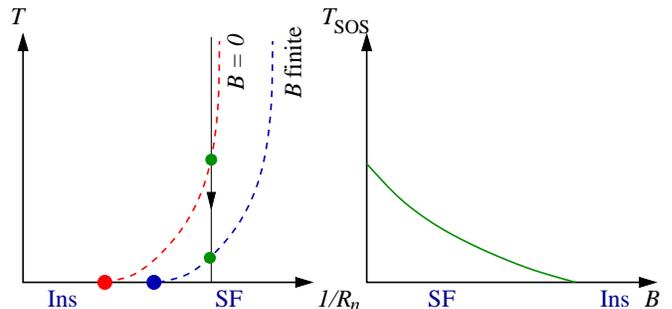} \caption{Finite
      magnetic field shifts the SOS crossover line (curved dashed
      lines). Consider cooling a system near the SI critical point at
      fixed $R_n$ (vertical line in left diagram). The
      crossover temperature $T_{\rm SOS}$ is indicated by the circle
      at the point where the vertical line intersects the SOS
      crossover line. This decreases with increasing $B$, as sketched
      in the diagram on the right.  Superfluid correlations develop in
      the system below $T_{\rm SOS}$. Above $T_{\rm SOS}$, the system
      may appear metallic.}
    \label{fig:bfield} 
  \end{center}
\end{figure}

How does this experimental result fit into our description? We
speculate that winding-number fluctuations are responsible for this
crossover behaviour. More specifically, the applied magnetic field
increases the winding number fluctuations in the system in the
representation where the bosons are Cooper pairs. (This can be viewed
as the bosonic analogue of positive magnetoresistance.) This moves the
the SOS transition/crossover line to lower values of $R_n$ (to the
right in Fig.\ \ref{fig:zerotemp}.) This is illustrated in Fig.\ 
\ref{fig:bfield}.  We see that, if the system is near the SI critical
point, the crossover temperature, $T_{\rm SOS}$, decreases rapidly
with increasing applied field $B$, as sketched in Fig.\ 
\ref{fig:bfield}. This may explain why the crossover from metal to
superconductor is only observed experimentally at low
applied fields\cite{e4}.

%


\section{Conclusions}

In this paper, we have revisited a model of dissipative Bose systems
where conventional theory \cite{chak86} predicts a direct
superfluid-insulator transition. By treating the phase fluctuations
more carefully, we developed a variational approach which can
distinguish between superfluid, normal and insulating phases. We can
confirm that a Bose-metal phase does not exist at zero temperature, in
agreement with conventional treatment. We are also able to establish
the regime of validity for the conventional treatment by studying
higher-order effects which couple the small and large phase
fluctuations.

We have argued that single-parameter scaling might break down because
of the existence of large phase fluctuations (the imaginary-time
``winding numbers'' of the order-parameter phase). There is a window
around the true critical point where strong winding-number
fluctuations persist down to exponentially low temperatures. This
means that superconductivity may not be observable in this regime at
experimentally accessible temperatures.  The width of this window of
crossover behaviour appears to be quite large in our mean-field
analysis. We expect that it would be renormalized in a more detailed
calculation, and that it would depend on details of the system (such
as the degree of disorder).

This window of crossover behaviour may be responsible for an apparent
metallic phase in some experiments\cite{e3}. The recent
observation\cite{e4} of an apparently metallic phase becoming
superconducting at very low temperatures appear to supports our
crossover picture near the superfluid-insulator critical point.

\begin{acknowledgments}
We would like to thank the British Council for financial support
(UK/HK Joint Research Scheme JRS98/40). DKKL is supported by the Royal
Society.
\end{acknowledgments}

\end{multicols}

\begin{references}
\bibitem{e1}  D.B. Haviland, Y. Liu and A.M. Goldman, \prl {\bf 62},
 2180(1989); see also A.M. Goldman and Y. Liu, Physica D {\bf 83},
 163(1995).
\bibitem{t1} M.P.A. Fisher, \prl {\bf 65}, 923(1990); M.P.A. Fisher,
 G. Grinstein and S.M. Girvin, \prl {\bf 64}, 587(1990).
\bibitem{t2} S. Chakravarty, S. Kivelson, G.T. Zimanyi and 
 B.I. Halperin, \prb {\bf 35}, 7256 (1987).
\bibitem{e2} H.M. Jaeger, D.B. Haviland, A.M. Goldman and B.G.
 Orr, \prb {\bf 34}, 4920 (1986).
\bibitem{e3} D. Ephron, A. Yazdani, A. Kapitulnik and B.R. Beasley,
 \prl {\bf 76}, 1529 (1996).
\bibitem{e4} N. Mason and A. Kapitulnik, cond-mat/00062138.

\bibitem{t3} M.V. Feigel'man and A.I. Larkin, cond-mat/9803006.
\bibitem{t4} D. Das and S. Doniach, cond-mat/9902308.
\bibitem{bs} J. Bardeen and M.J. Stephen, Phys. Rev. {\bf 140},
 A1197 (1965).
\bibitem{shim} E. Shimshoni, A. Auerbach and A. Kapitulnik, \prl
 {\bf 80}, 3352 (1998). 
\bibitem{ka} see N. Mason and A. Kapitulnik, cond-mat/9810228 for a
 similar idea.
\bibitem{chak86} S. Chakravarty, G. Ingold, S. Kivelson, and A. Luther,
\prl {\bf 56}, 2303 (1986).
\bibitem{fazio} see, for example, R. Fazio and G. Sch$\ddot{o}$n,
 \prb {\bf 43}, 5307 (1991); see also J.E. Mooij {\em et.al.},
 \prl {\bf 65}, 645 (1990).
\bibitem{aes} V. Ambegaokar, U. Eckern and G. Sch$\ddot{o}$n, \prl {\bf 48},
 1745 (1982).
\bibitem{cl} A.O. Caldeira and A.J. Leggett, Ann. Phys. (N.Y.) {\bf 149},
 374  (1983).
\bibitem{sz} see, for example, G. Sch$\ddot{o}$n and A.Z. Zaikin, \prb
 {\bf 40}, 5231 (1989).
\bibitem{dual} C. Dasgupta and B.I. Halperin, \prl {\bf 47}, 1556 (1981);
 M.P.A. Fisher and D.H. Lee, \prb{\bf 39}, 2756 (1989).
\end{references}
\end{document}